\documentclass[a4paper]{article}
\usepackage{INTERSPEECH2021}
\usepackage{physics} 
\usepackage{lipsum} 

\usepackage{hyperref}
\hypersetup{colorlinks=true, unicode=true, linkcolor=[rgb]{0.10,0.05,0.67}, citecolor=[rgb]{0.10,0.05,0.67}, filecolor=[rgb]{0.10,0.05,0.67}, urlcolor=[rgb]{0.10,0.05,0.67}}
\usepackage{multirow}
\usepackage{color, colortbl}
\definecolor{LightCyan}{rgb}{0.88,1,1}
\usepackage{xcolor}
\usepackage{caption}
\usepackage{subcaption}
\usepackage{amsmath}
\usepackage{float}
\usepackage{graphicx}
\usepackage{url}
\usepackage{hyperref}
\title{Exploring Disentanglement with Multilingual and Monolingual VQ-VAE}

\name{Jennifer Williams$^1$, Jason Fong$^1$, Erica Cooper$^2$, Junichi Yamagishi$^2$}
\address{
  $^1$Centre for Speech Technology Research, University of Edinburgh, UK\\
  $^2$National Institute for Informatics, Japan}
\email{$\{$j.williams,jason.fong$\}$@ed.ac.uk, $\{$ecooper,jyamagis$\}$@nii.ac.jp}

\begin{document}

\maketitle
\begin{abstract}
This work examines the content and usefulness of disentangled phone and speaker representations from two separately trained VQ-VAE systems: one trained on multilingual data and another trained on monolingual data. We explore the multi- and monolingual models using four small proof-of-concept tasks: copy-synthesis, voice transformation, linguistic code-switching, and content-based privacy masking. From these tasks, we reflect on how disentangled phone and speaker representations can be used to manipulate speech in a meaningful way. Our experiments demonstrate that the VQ representations are suitable for these tasks, including creating new voices by mixing speaker representations together. We also present our novel technique to conceal the content of targeted words within an utterance by manipulating phone VQ codes, while retaining speaker identity and intelligibility of surrounding words. Finally, we discuss recommendations for further increasing the viability of disentangled representations.
\end{abstract}


\section{Introduction}
One of the main benefits of using Vector Quantization Variational Autoencoders (VQ-VAE) for speech synthesis is that this architecture facilitates learning rich representations of speech \cite{zhao2020improved, williams2020learning, oord2017neural, yasuda2020end} in the form of discrete latent sequences. These learned representations come from vector-quantized \textit{codebooks} that behave as a clustering space with prototype centroids. Each entry in a codebook is represented by a pair consisting of a \textit{code} (also known as an index or token) and its corresponding vector. The code is a discrete integer value, and the vector is a learned \textit{n}-dimensional array of continuous values. In this paper, we are interested in the content and usefulness of codebooks after a VQ-VAE model has been trained. Specifically, we are the first to compare multilingual and monolingual VQ-VAE codebook representations for phone and speaker, with the aim to observe how well they adapt to voice transformation, linguistic code-switching and content-masking. 

The original VQ-VAE architecture design was based on a single VQ space: one encoder, one VQ codebook, and one decoder. That design proved to be useful across different objectives in image, video, and speech processing \cite{oord2017neural}. Since then, others have shown that the architecture could be expanded by stacking encoders which result in learning multiple different VQ spaces at the same time \cite{williams2020learning,zhang2019learning} or even hierarchical representations \cite{dhariwal2020jukebox}. These extended models provide more generalization capability, in part because they learn richer representations. 

It is possible to model multiple types of information in the speech signal with little or no supervision. In the process of learning to represent different types of information, the stacked VQ-VAE architectures are also providing a means to separate informational factors. This act of separating information from representations is known by several names, including factorization and disentanglement. Traditionally, factorization has served the purpose of removing irrelevant information from a representation such as a speaker embedding -- and then discarding what had been deemed irrelevant \cite{dehak2010front}. After information has been removed, it could be argued that a representation is in some way more ``pure''. On the other hand, disentanglement retains information. At the time of this writing we use the term \textit{disentanglement} to describe the phenomenon of isolating multiple types of distributed information from one source, into separate external representations. Functionally, this is a form of distributed representation learning. 

Currently there are no single-best techniques to measure the intrinsic goodness of disentangled representations apart from probing how well they perform in extrinsic tasks \cite{Raj_2019, peri2020empirical, williams2019disentangling, chung2020vector}. Recent efforts for phone and speaker disentanglement have been limited to contrastive tasks such as phone recognition and speaker recognition \cite{williams2020learning, ebbers2020contrastive}. Or observing that one representation ``gains'' information while another ``loses'' information \cite{williams2019disentangling, parthasarathi2012wordless} by measuring changes in classification accuracy.

Our work adds additional task-based evaluation by exploring disentanglement in both a multilingual and monolingual model. In order for the multilingual model to perform well at tasks such as voice transformation and linguistic code-switching, the learned representations must completely separate phonetic content and speaker information. We also introduce a novel technique that uses VQ phone codes to manipulate targeted content in the speech signal without altering the sound of a speaker's voice. Our exploration exposes some of the interesting capabilities of disentangled representations. We also offer ideas for improving the VQ-VAE architecture.

\section{Related Work}
Early versions of the VQ-VAE architecture with a single encoder and VQ phone codebook are known to be well-suited to voice conversion. Particularly \cite{ding2019group} showed that grouping latent embeddings together during the training process helps with mispronunciations. Their system relied on one-hot speaker encodings, but they suggest that the model could be made to generalize to unseen speakers by using externally-learned speaker embeddings instead. Our VQ-VAE implementation uses a similar approach to group latent embeddings, but goes one step further to simultaneously learn VQ speaker and phone embeddings.  

In \cite{wu2020one}, they propose a VQ technique that disentangles speaker and content information in a fully unsupervised manner for monolingual one-shot voice conversion. Phone embeddings originate from a VQ codebook whereas speaker embeddings are learned as a difference between discrete VQ codes and continuous VQ vectors. Finally, the speaker and content representations are re-combined additively (instead of by concatenation) and passed to the decoder as local conditions. While the method works very well in one-shot voice conversion, it does require a target speaker sample. Since the speaker representations rely on differences between internal VQ embeddings, it is not clear how the content and speaker representations could be used externally to this system, or whether or not it works for multilingual data. 

A dual-encoder VQ-VAE was proposed by \cite{zhao2020improved} which modeled the phone content and F0. This approach of using two encoders and learning two VQ codebooks was also used in \cite{williams2020learning} who sought to learn speaker identity as well as speech content at the same time. In \cite{williams2020learning}, they explored several variations of dual-encoder approach with different kinds of supervision. They found that the adversarial model performed disentanglement best between the speaker and content. In this paper, we utilize their pre-trained English VCTK model for multilingual adaptation as well as our experiments. 

While VQ-VAE has received a lot of attention for its potential in voice conversion, other challenges remain for multilingual speech synthesis. In \cite{himawan2020speaker} and \cite{zhou2019novel}, they showed it is possible to use DNNs to synthesize voices across languages, but these methods perform speaker adaptation rather than learning embeddings that could be re-purposed. Therefore these methods require an exemplar sentence that contains specific words and phrases. Likewise \cite{zhang2019learning, yang2020towards, li2019bytes} propose universal multi-language multi-speaker TTS systems, but it is not clear that the internal embeddings are re-useable for other speech tasks and the number of evaluated languages is small. 

Speech is often a primary medium for communicating sensitive information such as financial details or medical information. To date, most speech privacy scenarios reflect the need to protect speaker voice characteristics \cite{qian2018towards, tomashenko2020introducing}. The work of \cite{ahmed2020preech} proposes shuffling audio in a speech file to transform it into a speech ``bag of words'' so that the content and meaning cannot be easily gleaned from ASR. Likewise \cite{parthasarathi2012wordless} proposes using acoustic transformations to conceal the words of speech audio. Our approach to content privacy is inspired by \cite{hashimoto2016privacy} which created a \textit{speech privacy sound}. However, instead of privacy for speaker identity, we mask targeted words in a phrase by manipulating the sequence of discrete VQ phone codes.

\section{Data}
The multilingual SIWIS dataset \cite{goldman2016siwis} contains four languages: English, German, French, and Italian. There are 36 unique speakers. Each speaker is bilingual or trilingual and has been recorded in two or three languages. The dataset languages were imbalanced, so our train/test splits also preserved this imbalance as shown in Table~\ref{tab:data_splits}. The monolingual English VCTK dataset \cite{yamagishi2019cstr} contains 109 speakers with different accents. For VCTK, we used the same train/test splits as in \cite{williams2020learning}. All audio was downsampled to 16 kHz and normalized with sv56. The preprocessing steps were followed using scripts provided by \cite{zhao2020improved}.

\begin{table}[h]
\small
 \caption{SIWIS data splits across languages and speakers.}
 \centering
 \begin{tabular}{|l|cc|cc|cc|}
\hline
Language & \multicolumn{2}{c|}{Training} & \multicolumn{2}{c|}{Validation} & \multicolumn{2}{c|}{Held-out}\\
& Spk & Utt & Spk & Utt & Spk & Utt\\\hline\hline
English (EN) & 18 & 2387 & 18 & 603 & 4 & 16\\ \hline
French (FR) & 26 & 3405 & 26 & 841 & 5 & 16 \\ \hline
German (DE) & 13 & 1719 & 13 & 376 & 4 & 18 \\ \hline
Italian (IT) & 13 & 1689 & 13 & 430 & 3 & 10\\\hline
 \end{tabular}
 \label{tab:data_splits}
\end{table}




\begin{figure}
\centering
\includegraphics[width=0.45\textwidth]{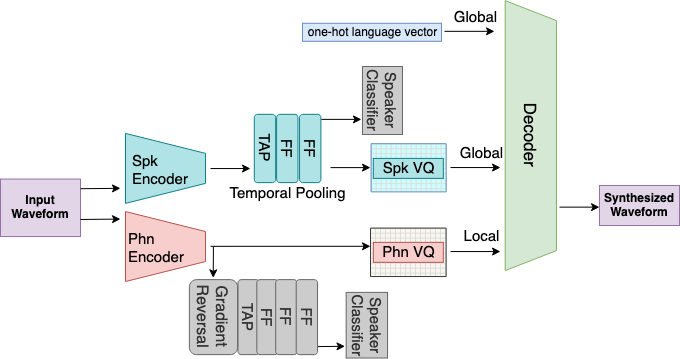} 
\caption{VQ-VAE overview from \cite{williams2020learning}, two encoders and VQ spaces which modeled speaker identity as a global condition, and speech phones as a local condition. We added a global one-hot language vector for our multilingual training.}
\label{fig:model}
\end{figure}

\section{VQ-VAE Model Adaptation}
We started with a dual-encoder VQ-VAE model that was pre-trained and provided by \cite{williams2020learning}. It learned two separate encoders and two separate VQ codebooks for speech content and speaker identity (Figure~\ref{fig:model}). They had trained the model to 500k steps using English VCTK data. 

We used the pre-trained model from \cite{williams2020learning} and adapted it to multilingual SIWIS data. For the model adaptation, a projection layer from the pre-trained WaveRNN decoder was discarded but we kept all other parameters from the encoders and VQ codebooks. We also added a one-hot language vector as global conditions to the WaveRNN decoder. We trained the multilingual model on all four languages mixed together for 550k steps while monitoring the validation losses. 

The goal is not to learn to disentangle languages, but to learn representations of content and speaker that are shared across multiple languages. For example, to learn phone VQ representations from multiple languages in a single VQ codebook. During the model adaptation, we did not experiment with changing the codebook sizes from the pre-trained model. Therefore we used a codebook size of 256 for the speaker codebook, and 512 for the phone codebook.

The input to the encoder was a waveform. After the waveform was downsampled by each encoder, it was transformed into a sequence of VQ codes and vectors for phones, and a single VQ code and vector for speaker identity. The VQ vectors were then provided to the WaveRNN decoder. Finally the output was a reconstructed waveform.

\section{Task-Based Evaluation}
The purpose of a task-based evaluation is to understand how learned phone or speaker representations perform in tasks that benefit from disentanglement. We describe four very small ``proof-of-concept'' tasks and corresponding results. The synthesized speech\footnote{Speech examples: \url{https://rhoposit.github.io/ssw11}} was assessed using human listening judgements. 
For the listening tests, participants were recruited from the Prolific\footnote{\url{https://www.prolific.co/}} platform and the listening test materials were hosted by Qualtrics\footnote{\url{https://www.qualtrics.com/uk/}}. 
We grouped our listening test tasks on the basis of language and dataset in order to utilize similar participants. This resulted in a total of seven separate listening tests and also allowed for consistency among our listener pool. For example, the same set of French speakers evaluated French MOS copy-synthesis, French MOS voice transformation, and French voice transformation speaker similarity. All of our participants self-identified as ``fluent'' in their respective languages, including pairs for code-switching: English-French, or English-German. While the multilingual model training included Italian data, this language was omitted from the evaluation as there were few speakers in the held-out set to select representative samples for gender, as well as bilingual/trilingual overlap. For each of the seven listening tests, we recruited 20 people and they were compensated at the rate of \pounds\ 7.50 per hour.

\subsection{Copy-Synthesis}
One way to gauge the quality of a trained VQ-VAE is to perform copy-synthesis. If copy-synthesis quality is very good then the internal VQ representations are more likely to also be good, however this is not guaranteed. While this does not inform us about the quality of the internal representations, it provides a starting point. This section is included as a sanity check. However, since the listening test was very small the reported MOS values may not generalize.

Listeners rated the naturalness on a Likert scale of 1-5 (where 5 is natural). We evaluated 6 examples per language using data from the held-out set, for a total of 24 samples. We report the average MOS naturalness scores in Table~\ref{tab:vqvae_quality}. The synthetic speech results in lower MOS scores for the monolingual and multilingual models. In the multilingual model, English and German naturalness was lower. The MOS for French had the smallest change from natural to synthetic. Evaluating with higher quantities of speech samples would provide a better perspective of the average MOS scores per language.  
\begin{table}
\centering
\small
 \caption{MOS naturalness scores for copy-synthesis. Results are reported for the multilingual model (SIWIS data) as well as the monolingual English model (VCTK data).}
 \begin{tabular}{|l|c|cc|}
\hline
Data & Natural & Synthetic & $\Delta$\\
\hline\hline
SIWIS-EN & 4.1 & 1.6 & $\downarrow$ 2.5 \\
SIWIS-FR & 3.4 & 2.9 & $\downarrow$ 0.5 \\
SIWIS-DE & 3.7 & 2.5 & $\downarrow$ 1.2 \\ \hline
VCTK-EN & 4.0 & 3.3 & $\downarrow$ 0.7 \\\hline
 \end{tabular}
 \label{tab:vqvae_quality}
\vspace{-4mm}
\end{table}

\subsection{Voice Transformation}
We present results from a \textit{voice transformation} task. We tried to change the speaker identity by replacing the speaker code to one of other codes obtained after the VQ-VAE optimization. Individual speaker codes do not always correspond to speakers included in the training dataset and hence this is not a conversion to specific identity of a target speaker. But, we would be able change the speaker identity by replacing the VQ speaker codes while keeping the VQ phone codes unchanged. For each model, we identified which VQ speaker codes had been learned during training. Neither of the two models utilized all of the possible speaker codebooks (the codebook size was 256 for both models), even though both models were trained with multi-speaker data. In the multilingual model (SIWIS), there were 11 VQ speaker codebooks utilized for 36 unique speakers. In the monolingual model (VCTK), there were 18 VQ speaker codebooks utilized for 110 unique speakers. Our VQ-VAE model under-estimated the number of speakers and seems to merge some speakers into one cluster.

\subsubsection{Single-Representation}
This version of voice transformation changes one single speaker VQ code at a time, without mixing or combining speaker codes. For the multilingual model, we selected one male and female speaker (\textbf{spk13}-male, \textbf{spk04}-female) from the SIWIS data and seen conditions. Then we extracted the VQ phone and speaker codes. We replaced their speaker codes with each of the 11 multilingual VQ speaker codes from the codebook. We used 2 utterances per speaker, per language for a total of 12 examples. For the one-hot language vector, we used the language from the source sentence. For the monolingual model and codebook, we followed the same approach selecting a male and female speaker from the VCTK data and seen conditions (\textbf{p229}-female-English, \textbf{p302}-male-Canadian). We selected 2 utterances for each speaker, for a total of 4 examples.

\subsubsection{Mixed-Representations}
This version of voice transformation mixes speaker VQ codes to create new voices, in a spirit similar to \textit{zero-shot} voice conversion. Ideally, this could be done using various combinations of VQ speaker codes and weighting them. In this work, we mixed two representations by calculating an unweighted mean between two VQ codebook vectors. In a vector space, the resulting representation is a new centroid that is equidistant between the paired vectors. We randomly paired VQ speaker codes for each model, and then mixed them. We synthesized the same source utterances as before. 

\begin{figure*}
\centering
\subfloat[Multilingual model]{\label{fig1:a}\includegraphics[width=0.23\linewidth]{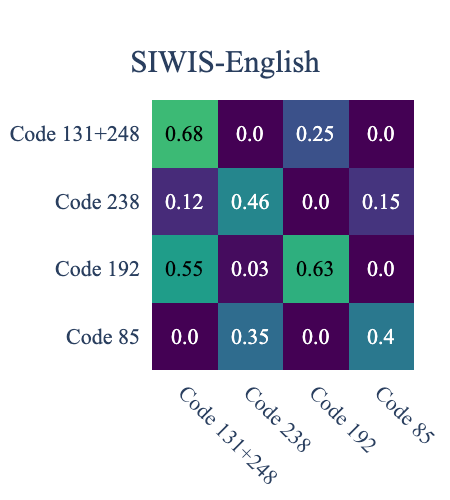}\includegraphics[width=0.165\textwidth]{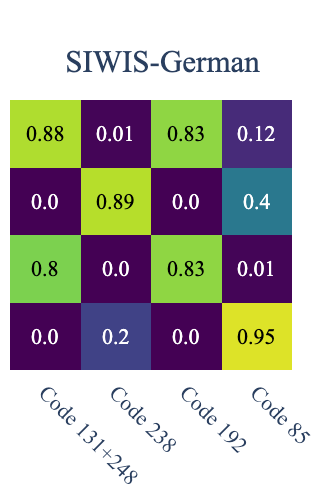}\includegraphics[width=0.19\textwidth]{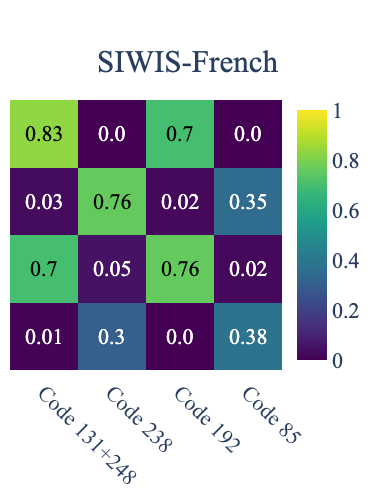}}\hfill
  \subfloat[Monolingual model]{\label{fig1:b}\includegraphics[width=0.25\textwidth]{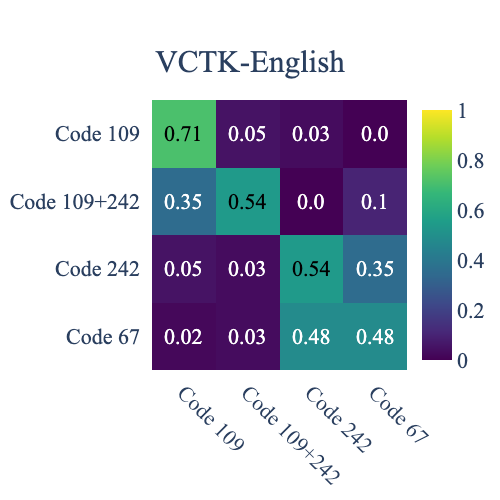}} \caption{Voice transformation speaker VQ code similarity matrix. Annotations represent the percent of listeners who marked a pair of utterances as the same speaker. Note that the monolingual and multilingual models utilize different speaker VQ codebooks. }
\label{fig:vc_similarity}
\end{figure*}

\begin{table}[ht]
\small
 \caption{Multilingual (SIWIS) MOS naturalness scores for voice transformation and voice mixing.}
 \centering
 \begin{tabular}{|l|c|c|c|}
\hline
Speaker Code & English & French & German\\\hline\hline
Code 85 & 2.4 & 2.9 & 3.4\\ 
Code 192 & 2.6 & 3.0 & 3.1 \\ 
Code 238 & 2.5 & 3.0 & 3.2 \\ \hline
Code 131+248 & 2.4 & 3.1 & 3.3\\\hline
 \end{tabular}
 \label{tab:vc_siwis_quality}
\end{table}

\begin{table}[h]
\small
 \caption{Monolingual English (VCTK) MOS naturalness scores for voice transformation and voice mixing.}
 \centering
 \begin{tabular}{|l|c|}
\hline
Speaker Code & English\\\hline\hline
Code 67 &  2.3\\ 
Code 109 & 2.3\\ 
Code 242 & 2.5\\ \hline
Code 109+242 & 2.4 \\\hline
 \end{tabular}
 \label{tab:vc_vctk_quality}
\vspace{-4mm}
\end{table}

\subsubsection{Results}
For the listening tests, we randomly selected 4 speaker VQ codes (3 single-representations, 1 mixed) from each model. Participants listened to all 8 samples in their language and marked naturalness on a scale of 1 to 5. The results for MOS naturalness are provided in Table~\ref{tab:vc_siwis_quality} and Table~\ref{tab:vc_vctk_quality}. MOS naturalness is changes depending on the speaker VQ code and language. The mixed VQ speaker vectors did not degrade the quality of the synthesized speech overall. In the multilingual model, French and German had better naturalness than English for all four of the reported VQ speaker codes. This is a similar pattern for naturalness in the earlier copy-synthesis task.

We also asked our listeners about speaker similarity. The purpose of this was to understand the consistency of the VQ speaker codes. Listeners were provided with matched and unmatched pairs in an A/B test, and were asked to decide if the A/B examples were from the same or different speaker. For example, a matched pair was 2 synthetic speech utterances using the target speaker VQ code \textbf{238}. An unmatched pair was 2 synthetic speech samples using two different speaker codes such as \textbf{238} and \textbf{85}. There were 16 total matched pairs and 24 unmatched pairs per language and dataset. This format allowed us to observe similarities and differences across a particular language and speaker VQ code. Recall that our voice transformation task did not utilize target speakers, only the learned VQ codes from the speaker codebooks. Speaker similarity results are reported in Figure~\ref{fig:vc_similarity}. The annotations in the figure represent the percent of listeners who marked a pair of utterances as the same speaker. A clear diagonal would indicate that the speaker VQ codes are consistently unique. In the multilingual model codes \textbf{131+248} and \textbf{192} are less consistent. German appears to be more consistent than French or English. In the monolingual model, we observed a pair of VQ speaker codes that participants identified as being inconsistent: \textbf{67} and \textbf{242}.

\begin{table}[ht!]
\small
 \caption{Speaker similarity for linguistic code-switching. \textit{A/B} measured how often listeners said the speaker was the same between synthetic and natural speech. \textit{Inter-Utt} measured how often listeners reported consistent speaker within an utterance. }
 \centering
 \begin{tabular}{|l|cc|}
\hline
& \multicolumn{2}{c|}{Speaker Similarity}\\
Data & A/B & Inter-Utt \\
\hline\hline
English-French & 57.9\% & 69.0\%\\
French-English & 30.8\% & 60.7\%\\\hline
English-German & 67.5\% & 77.5\%\\
German-English & 75.0\% & 77.5\%\\\hline

 \end{tabular}
 \label{tab:codeswitch}
\vspace{-4mm}
\end{table}

\subsection{Linguistic Code-Switching}
The purpose the linguistic code-switching task was to find out if we could generate speech using analysis-synthesis, wherein the speech has multiple languages within the same utterance. We simulated code-switching by concatenating together VQ phone codes from utterances in different languages but from the same speaker. This was possible because the SIWIS data contained utterances from bilingual and trilingual speakers. We used the sequence of VQ phone codes from entire audio files instead of word or phrase level granularity, and we did not change or modify the VQ phone code contents. We selected 6 utterances for English and German, and 6 utterances for English and French using both male and female speakers from the held-out set. We also swapped the language order, essentially doubling the number of exemplars. This was to observe if the WaveRNN decoder is sensitive to language ordering, since the decoder could only accept a single one-hot language code. This resulted in 24 code-switched files (6 per language and order pair). For the one-hot language vector, we used the language of the first utterance. The speech was synthesized from VQ phone and speaker codes without performing any modifications to the codes apart from the concatenation.

Our main interest for this task was to find out if the multilingual model could preserve speaker similarity while also synthesizing the multilingual speech. Listeners were presented with (A) code-switched synthetic speech from concatenated VQ phone codes, and (B) code-switched speech from concatenated audio files. In this A/B test, participants were asked if the speaker was the same between the two A/B samples. 

We also presented listeners with single code-switched examples from only (A) and asked the listeners to judge if the speaker voice was consistent throughout an utterance, or if it changed. This was measured because we had sometimes observed that the speaker voice was not consistent within an utterance. Results are reported in Table~\ref{tab:codeswitch}. We observed slightly more consistency for English-German pairs, compared to French. The A/B similarity for the French-English pair was particularly low, which means that the decoder had difficulty switching from French to English. This could be due to the language imbalances in the SIWIS dataset, or differences in the VQ phone code frequencies between these two languages. More investigation would uncover which part of the utterance was failing, and why the decoder was unable to recover. Better performance on German was also reflected in the other tasks. 

This analysis-synthesis task does not reflect how code-switching works with speakers in real-life because it was done at the utterance level instead of the word or phrase-levels. As mentioned earlier, the purpose was to observe if the model, especially WaveRNN, is capable of it. More investigation is required to understand and quantify the limits and edge cases of VQ-VAE for code-switching. In addition, the quantity of evaluated samples was particularly small, which makes it difficult to generalize the results or draw strong conclusions. We attempted to also measure intelligibility, however the listeners did not follow instructions often enough to perform calculations of intelligibility scores. For example, some listeners identified the names of the languages rather than the words of the utterance.

\subsection{Content-Based Privacy Masking}
The purpose of exploring content-based privacy is to develop a capability that conceals certain sensitive words or phrases in a manner that does not disrupt the normal flow and feel of a speech utterance. For example, in some use-cases it might be preferable to transform a sensitive phrase into a speaker's mumbling voice instead of a cut, beep, silence or static. Different types of masks may affect speech recognition (ASR) or speaker verification (ASV) differently. 

\begin{figure}
\centering
     \includegraphics[width=1.0\linewidth]{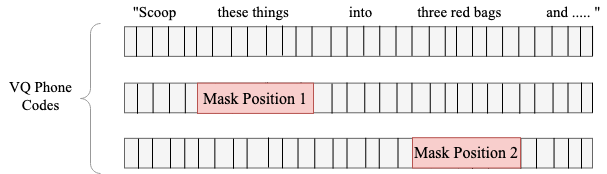}
     \caption{Diagram showing two different content masking positions for VQ phone codes on a given phrase.}
\label{fig:masking}
\end{figure}

In this task, we used the monolingual model because we had reliable alignments for the VCTK data \cite{mcauliffe2017montreal}. We hand-selected phrases that occurred mid-utterance and concealed them to try and render the target phrases unintelligible, while keeping the surrounding words intelligible. First, we used the forced-alignments to determine the timestamp end-points of the target phrase. Next, we used those endpoints to determine the location of the target phrase in the sequence of VQ phone codes. Finally, we modified only the VQ phone codes corresponding to the target phrase. We experimented with two different masking positions, as shown in Figure~\ref{fig:masking}, as well as two different masking methods. We have taken advantage of forced-alignments in this toy problem as well as knowing the target phrases beforehand. In real-world applications it may require keyword spotting or another mechanism to decide which words and phrases get masked. Performing this in real-time versus from a speech database would introduce additional engineering challenges. 

The first masking method was to replace true VQ phone codes of the target phrase with VQ phone codes from ICRA noise signals \cite{dreschler2001icra}.  Since the noise has speech-like spectral and temporal properties, it is expected to generate speech-like, but, meaningless phone codes. The speech-shaped noise offers a non-recoverable masking, which is useful for applications where speech content redaction must be persistent. First, we analyzed this noise to obtain its VQ phone codes. Even though the noise does not truly contain phones, the resulting VQ phone code sequence represented the noise quite well. Next, we replaced the sequence of true VQ phone codes for our target phrase with a randomly selected sequence of the SSN VQ codes of the same length. Our second technique was to simply reverse the the order of the true VQ phone codes for the target phrase, while leaving the remaining VQ phone codes intact. The VQ code reversal method does render the target phrase unintelligible, however it could be recovered by playing the audio backwards. We did not attempt other masking methods, however it may be possible to use silence or randomly selected VQ phone codes. It is also unknown if VQ-VAE could be used for recoverable masking, wherein the masked could be undone. Whether or not this is desirable depends on the use-case.

\subsubsection{Results}
We selected two utterances that were shared between a female and male speaker. Next, we selected two target phrases to mask, at different positions in the sentence. For the first utterance, the two target phrases were ``these things'' (position1) and ``three red bags'' (position2). For the second utterance, the two target phrases were ``sunlight strikes'' (position1) and ``raindrops in the air'' (position2). In total, 16 examples were evaluated. 

Participants were instructed that one or more words had been removed from the utterance, but were not told which ones. They were asked whether or not the speaker voice was consistent throughout the utterance and we measured the proportion of positive responses as shown in Table~\ref{tab:masking}. Overall the SSN was better for maintaining speaker identity throughout the utterance. In general, masking the phrase at position2 resulted in more consistency, which could be due to the challenges of using an auto-regressive decoder like WaveRNN. Listeners also performed an A/B preference test which revealed a slight preference for SSN over reversal masking. Finally, we measured ASR-based intelligibility as word error rate (WER) using the IBM Watson Speech-to-Text API\footnote{\url{https://www.ibm.com/cloud/watson-speech-to-text}}. We first calculated the WER on natural, unmasked audio as a baseline and found it was 24\%. This is higher than expected but likely due to pronunciations and the audio quality. The other WER is reported in Table~\ref{tab:masking}. Overall, the WER increased compared to natural, unmasked speech. The position1 resulted in better intelligibility, and the two different techniques were comparable on average. It is unclear if the rise in WER is due to the masking or if intelligibility was lost for unmasked words. Future work must provide a procedure to better evaluate content-based masking.

\begin{table}[h]
\small
 \caption{Speaker similarity and ASR-based WER for content masking, comparing two methods and target phrase positions.}
 \centering
 \begin{tabular}{|l|c|c|}
\hline
 & Speaker & ASR-Based\\
Masks & Similarity & WER\\
\hline\hline
Reversal Position1 & 63.7\% & 47\% \\\hline
Reversal Position2 & 77.5\% & 68\% \\\hline
SSN Position1 & 70.0\% & 53\% \\\hline
SSN Position2 & 76.2\% & 61\% \\\hline
 \end{tabular}
 \label{tab:masking}
\vspace{-4mm}
\end{table}

\section{Discussion}
We have shown that it is possible to adapt an existing monolingual VQ-VAE model to a new multi-speaker multi-language dataset with reasonable performance on copy-synthesis, voice transformation, and linguistic code-switching\footnote{Code/models: \url{https://github.com/rhoposit/multilingual_VQVAE}}. This is an important finding for multi-lingual speech synthesis. 

The manner in which the VQ speaker codebooks are under-utilized for both models has some implications for the limitations of the VQ-VAE architecture. It is sometimes referred to as \textit{codebook collapse} analogous to posterior collapse in VAE. We observed similar codebook collapse in our VQ phone codebooks as the VQ speaker codebooks. In both models, the phone codebook size was set to 256, however the multilingual model utilized 161 entries and the monolingual model utilized 170 entries. The quantity of utilized entries is far greater than the size of a requisite phone set -- even in the multilingual model. We examined the distribution of VQ phone codes for each language in the multilingual model and found that all four languages utilized similar codebooks with similar frequencies. 

The diversity of the learned codebooks should be improved. The size of codebooks must be pre-determined at the time of initializing the architecture. As we have shown, VQ-VAE models can be adapted to new datasets, but having hard-coded constraints (such as the codebook sizes) may be a limiting factor. Our recommendation is to develop a way to dynamically add or remove VQ codebooks during the training process. This would make it possible to learn only and all of the codebook vectors that matter. The true capabilities of VQ-VAE modeling are limited by its toolkit implementation: the nature of the tensor graph and how it is used in memory does not accommodate dynamic modeling to its fullest potential. 

We have described a method to synthesize high-quality speech in multiple languages (including code-switching) from a single multilingual model, based on learned representations. This will be useful for speech-to-speech translation, controllable speech synthesis, and data augmentation. In future work, we are interested in adding additional internal representations to the dual-encoder VQ-VAE model in an effort to perform further disentanglement of speech signal characteristics.

\section{Acknowledgements}
We sincerely thank Evelyn Williams at the University of Edinburgh for helping implement the listening tests. This work was partially supported by the EPSRC Centre for Doctoral Training in Data Science, funded by the UK Engineering and Physical Sciences Research Council (grant EP/L016427/1) and University of Edinburgh; and by a JST CREST Grant (JPMJCR18A6, VoicePersonae project), Japan. Some of the numerical calculations were carried out on the TSUBAME 3.0 supercomputer at the Tokyo Institute of Technology.

\bibliographystyle{IEEEtran}
\bibliography{mybib}

\begin{thebibliography}{10}
\providecommand{\url}[1]{#1}
\csname url@samestyle\endcsname
\providecommand{\newblock}{\relax}
\providecommand{\bibinfo}[2]{#2}
\providecommand{\BIBentrySTDinterwordspacing}{\spaceskip=0pt\relax}
\providecommand{\BIBentryALTinterwordstretchfactor}{4}
\providecommand{\BIBentryALTinterwordspacing}{\spaceskip=\fontdimen2\font plus
\BIBentryALTinterwordstretchfactor\fontdimen3\font minus
  \fontdimen4\font\relax}
\providecommand{\BIBforeignlanguage}[2]{{%
\expandafter\ifx\csname l@#1\endcsname\relax
\typeout{** WARNING: IEEEtran.bst: No hyphenation pattern has been}%
\typeout{** loaded for the language `#1'. Using the pattern for}%
\typeout{** the default language instead.}%
\else
\language=\csname l@#1\endcsname
\fi
#2}}
\providecommand{\BIBdecl}{\relax}
\BIBdecl

\bibitem{zhao2020improved}
Y.~Zhao, H.~Li, C.-I. Lai, J.~Williams, E.~Cooper, and J.~Yamagishi, ``Improved
  {P}rosody from {L}earned {F}0 {C}odebook {R}epresentations for {VQ-VAE}
  {S}peech {W}aveform {R}econstruction,'' \emph{INTERSPEECH}, 2020.

\bibitem{williams2020learning}
J.~Williams, Y.~Zhao, E.~Cooper, and J.~Yamagishi, ``Learning {D}isentangled
  {P}hone and {S}peaker {R}epresentations in a {S}emi-{S}upervised {VQ-VAE}
  {P}aradigm,'' \emph{ICASSP}, 2021.

\bibitem{oord2017neural}
A.~v.~d. Oord, O.~Vinyals, and K.~Kavukcuoglu, ``{N}eural {D}iscrete
  {R}epresentation {L}earning,'' \emph{Advances in Neural Information
  Processing Systems 30 (NIPS 2017)}, 2017.

\bibitem{yasuda2020end}
Y.~Yasuda, X.~Wang, and J.~Yamagishi, ``{E}nd-to-{E}nd {T}ext-to-{S}peech
  {U}sing {L}atent {D}uration {B}ased on {VQ-VAE},'' \emph{ICASSP}, 2021.

\bibitem{zhang2019learning}
Y.~Zhang, R.~J. Weiss, H.~Zen, Y.~Wu, Z.~Chen, R.~Skerry-Ryan, Y.~Jia,
  A.~Rosenberg, and B.~Ramabhadran, ``{L}earning to {S}peak {F}luently in a
  {F}oreign {L}anguage: {M}ultilingual {S}peech {S}ynthesis and
  {C}ross-{L}anguage {V}oice {C}loning,'' \emph{INTERSPEECH}, 2019.

\bibitem{dhariwal2020jukebox}
P.~Dhariwal, H.~Jun, C.~Payne, J.~W. Kim, A.~Radford, and I.~Sutskever,
  ``{J}ukebox: {A} {G}enerative {M}odel for {M}usic,'' \emph{arXiv preprint
  arXiv:2005.00341}, 2020.

\bibitem{dehak2010front}
N.~Dehak, P.~J. Kenny, R.~Dehak, P.~Dumouchel, and P.~Ouellet, ``{F}ront-{E}nd
  {F}actor {A}nalysis for {S}peaker {V}erification,'' \emph{IEEE Transactions
  on Audio, Speech, and Language Processing}, vol.~19, no.~4, pp. 788--798,
  2010.

\bibitem{Raj_2019}
D.~Raj, D.~Snyder, D.~Povey, and S.~Khudanpur, ``{P}robing the {I}nformation
  {E}ncoded in {X}-{V}ectors,'' \emph{2019 IEEE Automatic Speech Recognition
  and Understanding Workshop (ASRU)}, Dec 2019.

\bibitem{peri2020empirical}
R.~Peri, H.~Li, K.~Somandepalli, A.~Jati, and S.~Narayanan, ``An empirical
  analysis of information encoded in disentangled neural speaker
  representations,'' in \emph{Proc. Odyssey 2020 The Speaker and Language
  Recognition Workshop}, 2020, pp. 194--201.

\bibitem{williams2019disentangling}
J.~Williams and S.~King, ``Disentangling {S}tyle {F}actors {F}rom {S}peaker
  {R}epresentations.'' in \emph{INTERSPEECH}, 2019, pp. 3945--3949.

\bibitem{chung2020vector}
Y.-A. Chung, H.~Tang, and J.~Glass, ``Vector-quantized autoregressive
  predictive coding,'' \emph{Proc. Interspeech 2020}, pp. 3760--3764, 2020.

\bibitem{ebbers2020contrastive}
J.~Ebbers, M.~Kuhlmann, T.~Cord-Landwehr, and R.~Haeb-Umbach, ``Contrastive
  {P}redictive {C}oding {S}upported {F}actorized {V}ariational {A}utoencoder
  for {U}nsupervised {L}earning of {D}isentangled {S}peech {R}epresentations,''
  \emph{ICASSP}, 2021.

\bibitem{parthasarathi2012wordless}
S.~H.~K. Parthasarathi, H.~Bourlard, and D.~Gatica-Perez, ``{W}ordless
  {S}ounds: {R}obust {S}peaker {D}iarization {U}sing {P}rivacy-{P}reserving
  {A}udio {R}epresentations,'' \emph{IEEE Transactions on Audio, Speech, and
  Language Processing}, vol.~21, no.~1, pp. 85--98, 2012.

\bibitem{ding2019group}
S.~Ding and R.~Gutierrez-Osuna, ``Group {L}atent {E}mbedding for {V}ector
  {Q}uantized {V}ariational {A}utoencoder in {N}on-{P}arallel {V}oice
  {C}onversion.'' in \emph{INTERSPEECH}, 2019, pp. 724--728.

\bibitem{wu2020one}
D.-Y. Wu and H.-y. Lee, ``One-{S}hot {V}oice {C}onversion by {V}ector
  {Q}uantization,'' in \emph{ICASSP 2020-2020 IEEE International Conference on
  Acoustics, Speech and Signal Processing (ICASSP)}.\hskip 1em plus 0.5em minus
  0.4em\relax IEEE, 2020, pp. 7734--7738.

\bibitem{himawan2020speaker}
I.~Himawan, S.~Aryal, I.~Ouyang, S.~Kang, P.~Lanchantin, and S.~King,
  ``{S}peaker {A}daptation of a {M}ultilingual {A}coustic {M}odel for
  {C}ross-{L}anguage {S}ynthesis,'' in \emph{ICASSP 2020-2020 IEEE
  International Conference on Acoustics, Speech and Signal Processing
  (ICASSP)}.\hskip 1em plus 0.5em minus 0.4em\relax IEEE, 2020, pp. 7629--7633.

\bibitem{zhou2019novel}
X.~Zhou, H.~Che, X.~Wang, and L.~Xie, ``{A} {N}ovel {C}ross-{L}ingual {V}oice
  {C}loning {A}pproach with a {F}ew {T}ext-{F}ree {S}amples,'' \emph{arXiv
  preprint arXiv:1910.13276}, 2019.

\bibitem{yang2020towards}
J.~Yang and L.~He, ``{T}owards {U}niversal {T}ext-to-{S}peech,'' \emph{Proc.
  Interspeech 2020}, pp. 3171--3175, 2020.

\bibitem{li2019bytes}
B.~Li, Y.~Zhang, T.~Sainath, Y.~Wu, and W.~Chan, ``{B}ytes are all you need:
  {E}nd-to-{E}nd {M}ultilingual {S}peech {R}ecognition and {S}ynthesis with
  {B}ytes,'' in \emph{ICASSP 2019-2019 IEEE International Conference on
  Acoustics, Speech and Signal Processing (ICASSP)}.\hskip 1em plus 0.5em minus
  0.4em\relax IEEE, 2019, pp. 5621--5625.

\bibitem{qian2018towards}
J.~Qian, F.~Han, J.~Hou, C.~Zhang, Y.~Wang, and X.-Y. Li, ``{T}owards
  {P}rivacy-{P}reserving {S}peech {D}ata {P}ublishing,'' in \emph{IEEE INFOCOM
  2018-IEEE Conference on Computer Communications}.\hskip 1em plus 0.5em minus
  0.4em\relax IEEE, 2018, pp. 1079--1087.

\bibitem{tomashenko2020introducing}
N.~Tomashenko, B.~M.~L. Srivastava, X.~Wang, E.~Vincent, A.~Nautsch,
  J.~Yamagishi, N.~Evans, J.~Patino, J.-F. Bonastre, P.-G. No{\'e}
  \emph{et~al.}, ``{I}ntroducing the {V}oice{P}rivacy {I}nitiative,''
  \emph{INTERSPEECH}, 2020.

\bibitem{ahmed2020preech}
S.~Ahmed, A.~R. Chowdhury, K.~Fawaz, and P.~Ramanathan, ``{P}reech: {A}
  {S}ystem for {P}rivacy-{P}reserving {S}peech {T}ranscription,'' in \emph{29th
  {USENIX} Security Symposium (USENIX Security 20)}, 2020, pp. 2703--2720.

\bibitem{hashimoto2016privacy}
K.~Hashimoto, J.~Yamagishi, and I.~Echizen, ``Privacy-preserving sound to
  degrade automatic speaker verification performance,'' in \emph{2016 IEEE
  International Conference on Acoustics, Speech and Signal Processing
  (ICASSP)}.\hskip 1em plus 0.5em minus 0.4em\relax IEEE, 2016, pp. 5500--5504.

\bibitem{goldman2016siwis}
J.-P. Goldman, P.-E. Honnet, R.~Clark, P.~N. Garner, M.~Ivanova, A.~Lazaridis,
  H.~Liang, T.~Macedo, B.~Pfister, M.~S. Ribeiro \emph{et~al.}, ``The {SIWIS}
  {D}atabase: {A} {M}ultilingual {S}peech {D}atabase {W}ith {A}cted
  {E}mphasis,'' in \emph{Proceedings of Interspeech}, no. CONF, 2016.

\bibitem{yamagishi2019cstr}
J.~Yamagishi, C.~Veaux, K.~MacDonald \emph{et~al.}, ``{CSTR} {VCTK} {C}orpus:
  {E}nglish {M}ulti-{S}peaker {C}orpus for {CSTR} {V}oice {C}loning {T}oolkit
  (version 0.92),'' 2019.

\bibitem{mcauliffe2017montreal}
M.~McAuliffe, M.~Socolof, S.~Mihuc, M.~Wagner, and M.~Sonderegger, ``Montreal
  {F}orced {A}ligner: {T}rainable {T}ext-{S}peech {A}lignment {U}sing
  {K}aldi,'' \emph{Proc. Interspeech 2017}, 2017.

\bibitem{dreschler2001icra}
W.~A. Dreschler, H.~Verschuure, C.~Ludvigsen, and S.~Westermann,
  ``Icra{N}oises: {A}rtificial {N}oise {S}ignals with {S}peech-{L}ike
  {S}pectral and {T}emporal {P}roperties for {H}earing {I}nstrument
  {A}ssessment,'' \emph{Audiology}, vol.~40, no.~3, pp. 148--157, 2001.

\end{thebibliography}

\end{document}